# Pressure-induced superconductivity and topological quantum phase transitions in a quasi-one-dimensional topological insulator: Bi$_4$I$_4$

Yanpeng Qi[1], Wujun Shi[1,2], Peter Werner[3], Pavel G. Naumov[1,4], Walter Schnelle[1], Lei Wang[1,5], Kumari Gaurav Rana[3], Stuart Parkin[3], Sergiy A. Medvedev[1], Binghai Yan[1,2,6]*, Claudia Felser[1]*

[1]Max Planck Institute for Chemical Physics of Solids, 01187 Dresden, Germany.
[2]School of Physical Science and Technology, ShanghaiTech University, 200031, Shanghai, China.
[3]Max Planck Institute of Microstructure Physics, Weinberg 2, 06120 Halle, Germany.
[4]Shubnikov Institute of Crystallography of Federal Scientific Research Centre "Crystallography and Photonics" of Russian Academy of Sciences, 119333 Moscow, Russia
[5]Department of Power and Electrical Engineering, Northwest A&F University, 712100 Yangling, Shaanxi, China.
[6]Max Planck Institute for the Physics of Complex Systems, 01187 Dresden, Germany.

**Abstract**

**Superconductivity and topological quantum states are two frontier fields of research in modern condensed matter physics. The realization of superconductivity in topological materials is highly desired, however, superconductivity in such materials is typically limited to two- or three-dimensional materials and is far from being thoroughly investigated. In this work, we boost the electronic properties of the quasi-one-dimensional topological insulator bismuth iodide β-Bi$_4$I$_4$ by applying high pressure. Superconductivity is observed in β-Bi$_4$I$_4$ for pressures where the temperature dependence of the resistivity changes from a semiconducting-like**

**behavior to that of a normal metal. The superconducting transition temperature $T_c$ increases with applied pressure and reaches a maximum value of 6 K at 23 GPa, followed by a slow decrease. Our theoretical calculations suggest the presence of multiple pressure-induced topological quantum phase transitions as well as a structural-electronic instability.**

---

[*] E-mail: Claudia.Felser@cpfs.mpg.de; Yan@cpfs.mpg.de

**Introduction**

Dirac materials such as topological insulators (TI)[1, 2, 3], Dirac semimetals (DSM)[4, 5, 6, 7, 8, 9, 10, 11, 12], and Weyl semimetals (WSM)[13, 14, 15, 16, 17, 18, 19, 20, 21] have topologically nontrivial band structures and therefore exhibit unique quantum phenomena. Achieving a superconducting state, which is the state of quantum condensation of paired electrons, in topological materials has already led to some unprecedented discoveries. Indeed, the realization of superconductivity in topological compounds has been regarded as an important step toward topological superconductors.

Superconductivity has been induced by using doping or pressure in TIs ($Bi_2Se_3$[22, 23, 24, 25], $Bi_2Te_3$[26, 27, 28], $Sb_2Te_3$[29]), DSMs ($Cd_3As_2$[30, 31, 32], $ZrTe_5$[33] and $HfTe_5$[34]) and WSMs ($TaAs$[35], $TaP$[36], $WTe_2$[37, 38] and $MoTe_2$[39]). However, from a structural perspective, most topological materials are limited to two- or three-dimensional structures. Superconductivity has not been thoroughly explored in low-dimensional topological materials. Recently $\beta$-$Bi_4I_4$ has been theoretically predicted and experimentally confirmed as a new $Z_2$ TI[40]. Importantly, $\beta$-$Bi_4I_4$ crystallizes in a quasi-one-dimensional (quasi-1D) structure and thus hosts highly anisotropic surface-state Dirac fermions.

In this work, we systematically investigate the high-pressure behavior of the novel quasi-1D TI $\beta$-$Bi_4I_4$. Through ab-initio band structure calculations, we find that the application of pressure alters the electronic properties and leads to multiple topological quantum phase transitions: from strong TI (STI) to weak TI (WTI) and back to STI. Corresponding anomalies are visible in pressure-dependent resistivity data. Superconductivity is observed in $\beta$-$Bi_4I_4$ when the temperature dependence of $\rho(T)$ changes from a semiconducting–like behavior to that of a normal metal. The superconducting transition temperature $T_c$ increases with applied pressure and reaches a maximum value of 6 K at 23 GPa for β-Bi$_4$I$_4$, followed by a slow decrease.

## Results

**Structure and transport properties under ambient pressure.** Prior physical property measurements, β-Bi$_4$I$_4$ crystals used for the study were structurally characterized using single-crystal x-ray diffraction (SXRD) and high-angle annular dark-field scanning transmission electron microscopy (HAADF-STEM). Energy-dispersive x-ray spectroscopy (EDXS) analysis confirms that the single crystals are homogeneous and that the atomic ratio of elements is Bi:I = 53.8(2):46.2(4), in agreement with previously reported data[40]. β-Bi$_4$I$_4$ crystallizes in a monoclinic structure (space group $C12/m1$, No. 12), as shown in Fig. 1a, b. The 1D building blocks of β-Bi$_4$I$_4$, aligned along the $b$ axis, can be viewed as narrow nanoribbons of a bismuth bilayer (four Bi atoms in width) terminated by iodine atoms. The atomic arrangement of β-Bi$_4$I$_4$ was determined using HAADF-STEM images and diffraction patterns (Fig. 1c). One primitive cell consists of four I atoms and four Bi atoms which can be divided into two non-equivalent types of atoms: inner Bi1 atoms that bind to three bismuth atoms and peripheral Bi2 atoms that are saturated by covalent bonds to four iodine atoms.

**Electrical resistivity at high pressure.** In Figure 2 the temperature dependence of the resistivity $\rho(T)$ of β-Bi$_4$I$_4$ for various pressures is shown. For $P = 0.5$ GPa, $\rho(T)$ displays a semiconducting-like behavior similar to that observed at ambient pressure[40, 41], however our crystals do not show an upturn below $\approx 100$ K.[40] In a low-pressure region, increasing the pressure initially induces a weak but continuous suppression of the overall magnitude of $\rho$ with a minimum occurring at $P_{min} = 3$ GPa. Upon further increasing the pressure, the resistivity starts to increase gradually, reaching a maximum at a pressure above 8 GPa.

As the pressure is further increased above 8.8 GPa, $\rho$ rapidly decreases, exhibiting semiconductor-like behavior for β-Bi$_4$I$_4$ (Fig. 2b). As pressure increases up to 13.5 GPa, the normal state behaves as a metal, and a small drop of $\rho$ is observed at the lowest temperatures (experimental $T_{min}$ = 1.9 K). Zero resistivity is achieved for $P \geq$ 17.6 GPa, indicating the emergence of superconductivity. The critical temperature of superconductivity, $T_c$, gradually increases with pressure, and the maximum $T_c$ of 6 K is attained at $P$ = 23 GPa, as shown in Fig. 2c. Beyond this pressure, $T_c$ decreases slowly, showing a dome-like behavior (Fig. 2d).

The appearance of bulk superconductivity in β-Bi$_4$I$_4$ is further supported by the evolution of the resistivity-temperature curve with an applied magnetic field. The superconducting transition gradually shifts toward lower $T$ with the increase of the magnetic field (Fig. 2e). A magnetic field $\mu_0H$ = 2.5 T removes all signs of superconductivity above 1.9 K. The upper critical field $\mu_0H_{c2}$ is determined using the 90% points on the transition curves, and plots of $H_{c2}(T)$ are shown in Fig. 2f. A simple estimate using the conventional one-band Werthamer–Helfand–Hohenberg (WHH) approximation, neglecting the Pauli spin-paramagnetism effect and spin-orbit interaction[42], i.e., $\mu_0H_{c2}(0)$ = -0.693 × $\mu_0(dH_{c2}/dT) \times T_c$, yields a value of 2.5 T for β-Bi$_4$I$_4$. We also used the Ginzburg–Landau formula to fit the data:

$$H_{c2}(T) = H_{c2}(0)\frac{1-t^2}{1+t^2} \tag{1}$$

where $t$ = $T/T_c$, yielding a critical field $\mu_0H_{c2}$ = 2.7 T for β-Bi$_4$I$_4$. Both values are comparable with those determined for superconducting Bi$_2$Se$_3$ and BiTeI under pressure[24, 25, 43]. According to the relationship $\mu_0 H_{c2} = \Phi_0/(2\pi\xi^2)$, where $\Phi_0 = 2.07\times10^{-15}$ Wb is the flux quantum, the coherence length $\xi_{GL}(0)$ is 11.5 nm for β-Bi$_4$I$_4$. Note that the extrapolated values of $H_{c2}(0)$ are well below the Pauli–Clogston limit.

**Discussion**

The pressure dependence of the resistivity at room temperature and the critical temperature of superconductivity for β-Bi$_4$I$_4$ are summarized in Fig. 3. The resistivity of β-Bi$_4$I$_4$ exhibits a non-monotonic evolution with increasing pressure. Over the whole temperature range, the resistivity is first suppressed with applied pressure and reaches a minimum value at about 3 GPa. As the pressure further increases, the resistivity increases with a maximum occurring at 8 GPa. Then, the resistivity abruptly decreases. Superconductivity is observed after the temperature dependence of $\rho(T)$ changes from a semiconducting–like behavior to that of a metal. The superconducting $T_c$ increases with applied pressure, and a typical domelike evolution is obtained.

The presented results demonstrate that high pressure dramatically alters the electronic properties in β-Bi$_4$I$_4$. To obtain a comprehensive understanding of the physical properties of β-Bi$_4$I$_4$, we performed density-functional theory (DFT) calculations for the electronic band structures. Because of the underestimated band gap within the local density approximation or generalized gradient approximation, we employed the hybrid-functional method (HSE) to calculate the electronic properties. The calculated band structures and density of states (DOS) are displayed in Fig. 4 and Supplementary Figures 1 and 2. At zero pressure, the HSE calculations predict a narrow indirect band gap of 40.9 meV, with a valence band maximum (VBM) at the M point and a conduction band minimum (CBM) at the Y point. The component of VBM is mainly the *p* orbital of Bi1 with odd parity, while the CBM is mainly the *p* orbital of Bi2 with even parity. The band dispersion is relatively weak along the AΓYM path, which is perpendicular to the quasi-1D chain, indicating weak interaction between the chains. In contrast, the strong dispersion along the BΓ direction indicates strong interaction within the chain. Thus, the dispersion clearly reflects the quasi-1D character of β-Bi$_4$I$_4$.

From the band structure at zero pressure β-Bi$_4$I$_4$ is in a STI phase with a band inversion at the Y point. As pressure increases, the CBM and the VBM meet at the M point. Band inversion occurs and the structure is driven into a WTI phase[44]. For the process of band inversion, the band gap decreases to zero and then reopens. Therefore, the resistivity decreases before the band gap closes and then increases after the band gap reopens. This trend is roughly consistent with the experimental resistivity values (see Fig. 3). When pressure continues increasing, the band at the Y point is inverted back, and the structure returns to the STI phase. This phase evolution is also shown in Fig. 3b. When the pressure increases, the DOS near the Fermi level increases (see Fig. 4f). We also note that the increase of the DOS is mainly due to the flat bands near the Fermi level in the band structure. These heavy bands may exhibit low mobility, which may be the reason for the additional increase in resistivity in our experiments.

The pressure-induced multiple topological quantum phase transitions in β-Bi$_4$I$_4$ are unusual, and in addition β-Bi$_4$I$_4$ shows an electronic instability. DFT calculations indicate that the crystal structure abruptly changes at a critical pressure of 11.5 GPa. We can see that the lattice parameter along the quasi-1D chain direction decreases, while the parameters in the other two directions suddenly increase (Supplementary Fig. 3a,b). We also calculated the bond length within the Bi plane. Bond 2 (bond 1) suddenly increases (decrease) at the critical pressure (Supplementary Fig. 3c), which is further confirmed by the phonon spectrum (Supplementary Fig. 4). Near 11.5 GPa, an imaginary phonon mode appears, which corresponds to vibrations along the quasi-1D chain and leads to the collapse of the lattice along the chain direction. From the electronic band structure calculations we can see that, after the lattice constant changes, the structure is driven from an STI to a metal. The Fermi level crosses the band (see Fig. 4 e) and the DOS increases near the Fermi level (see Fig. 4f). Indeed, the resistivity abruptly decreases above the critical pressure and superconductivity is observed in β-Bi$_4$I$_4$ when the temperature dependence of $\rho(T)$ changes from a semiconducting–like behavior to that of a metal.

As a novel topological insulator, β-Bi$_4$I$_4$ offers a new platform for exploring exotic physics with simple chemistry. We find multiple topological quantum phase transitions under high pressure and β-Bi$_4$I$_4$ shows electronic instabilities. Superconductivity is induced after the nonmetal–to–metal transition in β-Bi$_4$I$_4$, which may be attributed to electronic and structure instabilities.

## Methods

**Single–crystal growth and characterization.** Single crystals of β-Bi$_4$I$_4$ were obtained from gas phase reactions using methods similar to those described in Refs[40, 41, 45]. Thoroughly ground mixtures of bismuth metal and HgI$_2$ were used as starting materials. The Bi to HgI$_2$ molar ratio was 1:2 with a total mass of ≈ 3 g. After evacuation and sealing, the ampoule was inserted into a furnace with a temperature gradient of 210 to 250°C with the educts in the hot zone. The ampoule was tilted by 20-30°, the cold end pointing upwards. After two weeks, needle-like crystals of size 5 × 1 × 0.5 mm had grown in the cold zone. The structures of the β-Bi$_4$I$_4$ crystals were investigated using single-crystal x-ray diffraction (SXRD) with Mo $K_a$ radiation. To analyze the atomic structure of the material, transmission electron microscopy (TEM) was performed.

**Experimental details of high-pressure measurements.** Resistivity measurements were performed under high pressure in a non-magnetic diamond anvil cell. A mixture of epoxy and fine cubic boron nitride (cBN) powder was used for the insulating gaskets, and platinum (Pt) foil with a thickness of 5 μm was used for electrodes. The diameters of the flat working surface of the diamond anvil and the hole in the gasket were 500 and 200 μm, respectively. The sample chamber thickness was ≈ 40 μm. Resistivity was measured using an inverting dc current with the van der Pauw technique implemented in a typical cryogenic setup at zero magnetic field, and the magnetic-field measurements were performed on a magnet-cryostat (PPMS-9, Quantum Design, $T_{min}$ = 1.8 K). Pressure was measured using the ruby scale for small chips of ruby placed in contact with the sample[46].

**Density functional theory calculations.** Density functional theory (DFT) calculations were performed using the Vienna Ab-initio Simulation Package (VASP)[47] with a plane-wave basis. The interactions between the valence electrons and ion cores were described using the projector-augmented wave method[48, 49]. The exchange and correlation energy was formulated using the generalized gradient approximation (GGA) with the Perdew–Burke–Ernzerhof scheme[50]. Van der Waals corrections were also included via a pairwise force field of the Grimme method[51, 52]. Because GGA usually underestimates the band gap, we used Heyd–Scuseria–Ernzerhof (HSE) screened Coulomb hybrid density functionals to calculate the electronic band structures and $Z_2$ topological invariant[53, 54]. The HSE band structure was obtained by the interpolated Wannier function supplied by the Wannier90 code[55]. The $Z_2$ topological invariant was calculated by the products of parity eigenvalues of all the occupied bands at the time-reversal-invariant momentum (TRIM) points[56]. The plane-wave basis cutoff energy was set to 176 eV by default. The $\Gamma$-centered $k$ points with 0.03 Å$^{-1}$ spacing were used for the first Brillouin-zone sampling. The structures were optimized until the forces on the atoms were less than 5 meV Å$^{-1}$. The pressure was derived by fitting the total energy dependence on the volume using the Murnaghan equation[57]. Note that spin-orbit coupling was included in the static calculation. The phonon dispersion was performed using the finite displacement method with VASP and PHOHOPY code[58], and a supercell with all lattice constants larger than 10.0 Å was employed to calculate the phonon spectra.

**Acknowledgments**

Y. Qi acknowledges financial support from the Alexander von Humboldt Foundation. The authors thank Horst Blumtritt for the FIB sample preparation and Dr. Yurii Prots for single-crystal x-ray diffraction studies. This work was financially supported by Deutsche Forschungsgemeinschaft (DFG; Project EB 518/1-1 of DFG-SPP 1666 "Topological Insulators") and by the European Research Council (ERC Advanced Grant No. 291472 "Idea Heusler").

**Figure Captions**

**Figure 1. Crystal structures of β-Bi$_4$I$_4$.** (**a**) Crystal structure of β-Bi$_4$I$_4$ viewed along the chain direction (lattice vector $b$ = [010] projection). The red, green, and blue spheres represent the Bi1, Bi2, and I atoms, respectively. Solid lines indicate covalent bonds. (**b**) Atomic structure of an individual chain-like building block of β-Bi$_4$I$_4$. (**c**) HAADF-STEM image of β-Bi$_4$I$_4$ along the [010] zone and corresponding electron diffraction pattern.

**Figure 2. Evolution of the electrical resistivity as a function of pressure for β-Bi$_4$I$_4$.** (**a,b**) Electrical resistivity as a function of temperature for pressures up to 40 GPa. The temperature dependence of $\rho(T)$ changes from semiconducting–like behavior to that of a normal metal. (**c,d**) Electrical resistivity drop and zero-resistivity behavior at low temperature for pressures of 13.5–23.0 and 24.1–39.5 GPa, respectively. Here, $T_c$ increases under increasing pressure and the maximum superconducting transition temperature corresponding to $T_c$ = 6 K at 23.0 GPa. (**e**) Temperature dependence of resistivity under different magnetic fields for β-Bi$_4$I$_4$ at 21.8 GPa. (**f**) Temperature dependence of the upper critical field for β-Bi$_4$I$_4$. Here, $T_c$ is determined as the 90% drop of the normal state resistivity. The solid lines represent the fits based on the Ginzburg–Landau formula.

**Figure 3. Electronic *P-T* phase diagram for β-Bi$_4$I$_4$.** (**a**) Superconducting $T_c$ is shown as function of pressure. The green and blue circles represent the $T_c$ extracted from different resistivity measurments. As the pressure increases to ≈ 13 GPa, superconductivity is observed above experimental $T_{min}$ and persists up to pressures near 40 GPa. A domelike evolution is observed with a maximum $T_c$ of 6 K attained at $P$ =

23 GPa. The resistivity values at 300 K are also shown. The resistivity of β-Bi$_4$I$_4$ exhibits a more complicated feature, which demonstrates that high pressure dramatically alters the electronic properties in β-Bi$_4$I$_4$. (**b**) Schematic illustrations of the band structure evolution under various pressures. In the STI phase, band inversion occurs at one time-reversal-invariant momentum (TRIM) points of the Brillouin zone; however, in the WTI phase, band inversion occurs at two TRIM points. In metals the Fermi level crosses the bands.

**Figure 4. Calculated band structure and density of state of β-Bi$_4$I$_4$.** (**a**)–(**e**) Band structure of β-Bi$_4$I$_4$ at 0 GPa, 2.4 GPa, 6.0 GPa, 11.1 GPa and 13.3 GPa**,** respectively. The size of red (blue) filled circles represents the fraction of Bi1 (Bi2) *5p* states. Band inversion clearly occurs between the Bi1 *5p* and Bi2 *5p* states. The dashed line represents the Fermi level. (**f**) Evolution of the electronic density of states with increasing pressure.

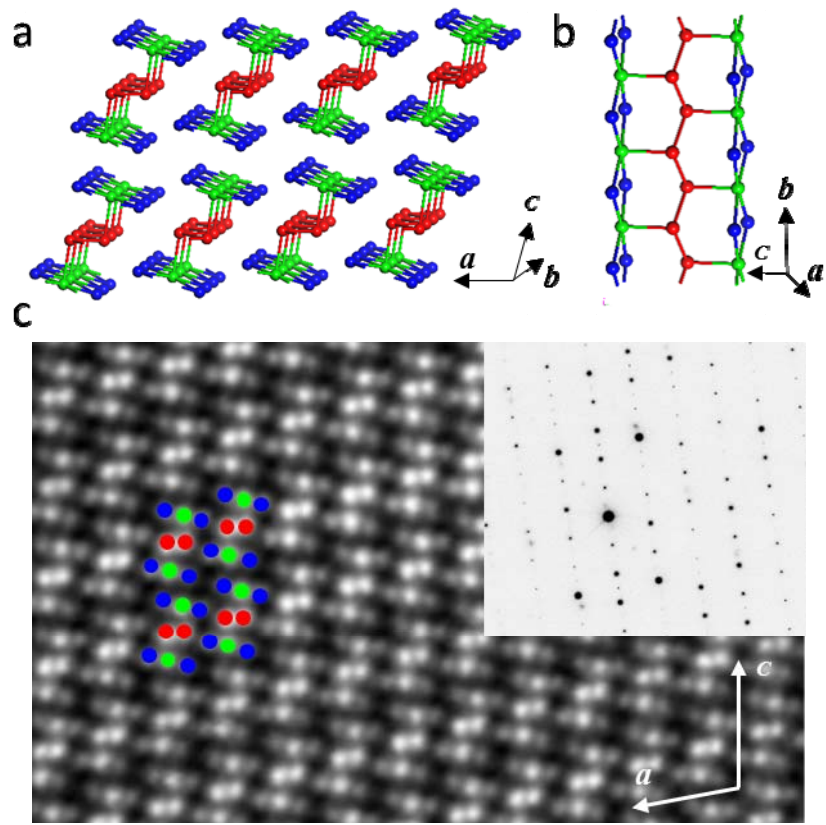

Fig. 1 Qi et al.

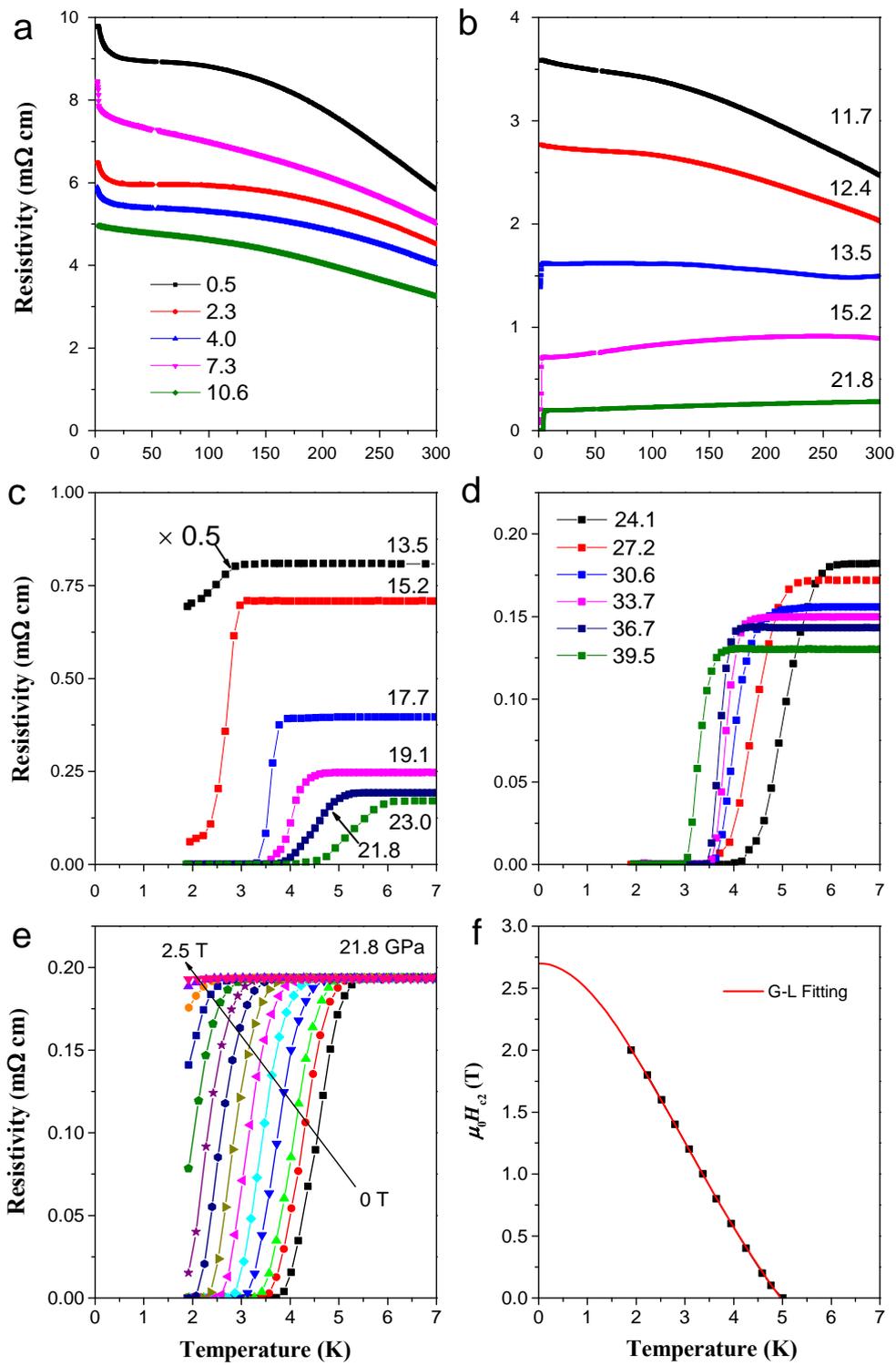

Fig. 2 Qi et al

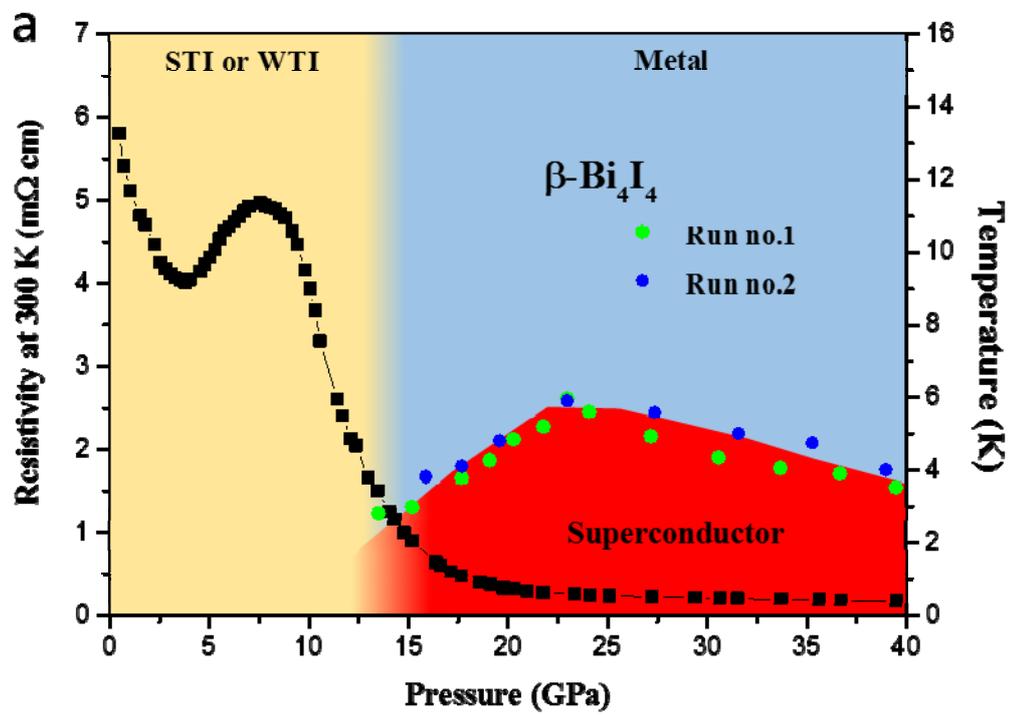

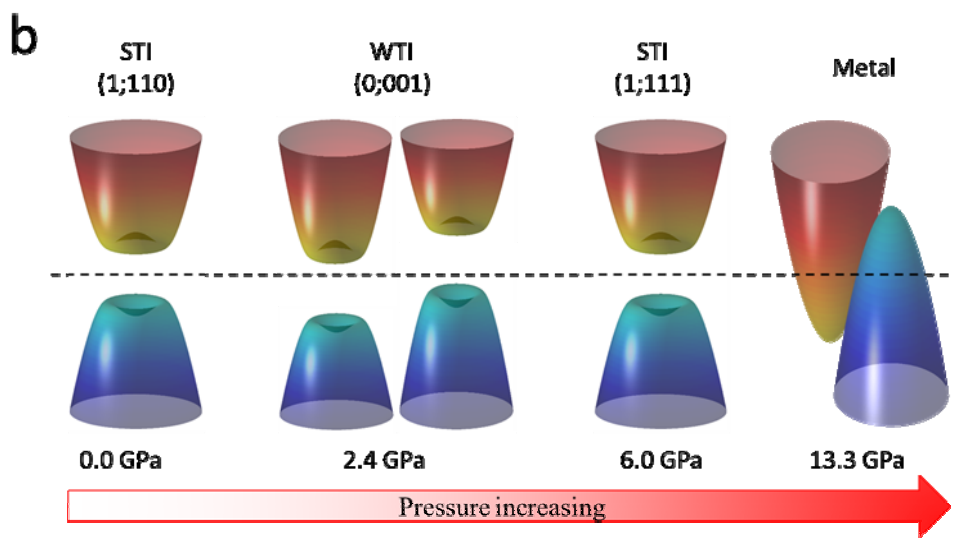

Fig. 3 Qi et al

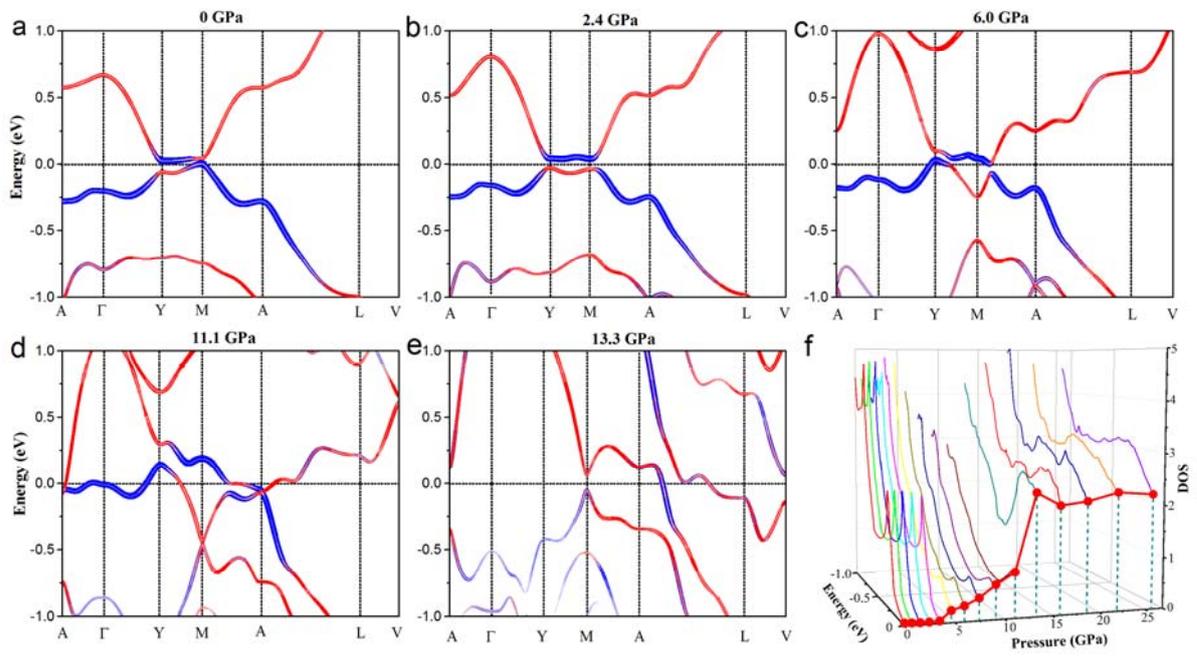

Fig. 4 Qi et al

# References

1. Fu, L., Kane, C. L., Mele, E. J. Topological Insulators in Three Dimensions. *Phys. Rev. Lett.* **98**, 106803 (2007).

2. Hasan, M. Z., Kane, C. L. Topological insulators. *Rev. Mod. Phys.* **82**, 3045-3067 (2010).

3. Qi, X.-L., Zhang, S.-C. Topological insulators and superconductors. *Rev. Mod. Phys.* **83**, 1057-1110 (2011).

4. Young, S. M., Zaheer, S., Teo, J. C. Y., Kane, C. L., Mele, E. J., Rappe, A. M. Dirac Semimetal in Three Dimensions. *Phys. Rev. Lett.* **108**, 140405 (2012).

5. Wang, Z., *et al.* Dirac semimetal and topological phase transitions in $A_3$Bi ($A$=Na, K, Rb). *Phys. Rev. B* **85**, 195320 (2012).

6. Wang, Z., Weng, H., Wu, Q., Dai, X., Fang, Z. Three-dimensional Dirac semimetal and quantum transport in $Cd_3As_2$. *Phys. Rev. B* **88**, 125427 (2013).

7. Liu, Z. K., *et al.* Discovery of a Three-Dimensional Topological Dirac Semimetal, $Na_3Bi$. *Science* **343**, 864-867 (2014).

8. Liu, Z. K., *et al.* A stable three-dimensional topological Dirac semimetal $Cd_3As_2$. *Nat. Mater.* **13**, 677-681 (2014).

9. Borisenko, S., Gibson, Q., Evtushinsky, D., Zabolotnyy, V., Büchner, B., Cava, R. J. Experimental Realization of a Three-Dimensional Dirac Semimetal. *Phys. Rev. Lett.* **113**, 027603 (2014).

10. He, L. P., *et al.* Quantum Transport Evidence for the Three-Dimensional Dirac Semimetal Phase in $Cd_3As_2$. *Phys. Rev. Lett.* **113**, 246402 (2014).

11. Xu, S.-Y., *et al.* Observation of Fermi arc surface states in a topological metal. *Science* **347**, 294-298 (2015).

12. Liang, T., Gibson, Q., Ali, M. N., Liu, M., Cava, R. J., Ong, N. P. Ultrahigh mobility

and giant magnetoresistance in the Dirac semimetal $Cd_3As_2$. *Nat. Mater.* **14**, 280-284 (2015).

13. Wan, X., Turner, A. M., Vishwanath, A., Savrasov, S. Y. Topological semimetal and Fermi-arc surface states in the electronic structure of pyrochlore iridates. *Phys. Rev. B* **83**, 205101 (2011).

14. Wan, X., Vishwanath, A., Savrasov, S. Y. Computational Design of Axion Insulators Based on $5d$ Spinel Compounds. *Phys. Rev. Lett.* **108**, 146601 (2012).

15. Burkov, A. A., Balents, L. Weyl Semimetal in a Topological Insulator Multilayer. *Phys. Rev. Lett.* **107**, 127205 (2011).

16. Xu, G., Weng, H., Wang, Z., Dai, X., Fang, Z. Chern Semimetal and the Quantized Anomalous Hall Effect in $HgCr_2Se_4$. *Phys. Rev. Lett.* **107**, 186806 (2011).

17. Bulmash, D., Liu, C.-X., Qi, X.-L. Prediction of a Weyl semimetal in $Hg_{1-x-y}Cd_xMn_yTe$. *Phys. Rev. B* **89**, 081106 (2014).

18. Weng, H., Fang, C., Fang, Z., Bernevig, B. A., Dai, X. Weyl Semimetal Phase in Noncentrosymmetric Transition-Metal Monophosphides. *Phys. Rev. X* **5**, 011029 (2015).

19. Huang, S.-M., *et al.* A Weyl Fermion semimetal with surface Fermi arcs in the transition metal monopnictide TaAs class. *Nat. Commun.* **6**, 7373 (2015).

20. Soluyanov, A. A., *et al.* Type-II Weyl semimetals. *Nature* **527**, 495-498 (2015).

21. Sun, Y., Wu, S.-C., Ali, M. N., Felser, C., Yan, B. Prediction of Weyl semimetal in orthorhombic $MoTe_2$. *Phys. Rev. B* **92**, 161107 (2015).

22. Hor, Y. S., *et al.* Superconductivity in $Cu_xBi_2Se_3$ and its Implications for Pairing in the Undoped Topological Insulator. *Phys. Rev. Lett.* **104**, 057001 (2010).

23. Kriener, M., Segawa, K., Ren, Z., Sasaki, S., Ando, Y. Bulk Superconducting Phase with a Full Energy Gap in the Doped Topological Insulator $Cu_xBi_2Se_3$. *Phys. Rev.*

*Lett.* **106**, 127004 (2011).

24.    Kirshenbaum, K., *et al.* Pressure-Induced Unconventional Superconducting Phase in the Topological Insulator $Bi_2Se_3$. *Phys. Rev. Lett.* **111**, 087001 (2013).

25.    Kong, P. P., *et al.* Superconductivity of the topological insulator $Bi_2Se_3$ at high pressure. *J. Phys.: Condens. Matter* **25**, 362204 (2013).

26.    Zhang, J. L., *et al.* Pressure-induced superconductivity in topological parent compound $Bi_2Te_3$. *Proc. Natl. Acad. Sci. U.S.A.* **108**, 24-28 (2011).

27.    Zhang, C., *et al.* Phase diagram of a pressure-induced superconducting state and its relation to the Hall coefficient of $Bi_2Te_3$ single crystals. *Phys. Rev. B* **83**, 140504 (2011).

28.    Matsubayashi, K., Terai, T., Zhou, J. S., Uwatoko, Y. Superconductivity in the topological insulator $Bi_2Te_3$ under hydrostatic pressure. *Phys. Rev. B* **90**, 125126 (2014).

29.    Zhu, J., *et al.* Superconductivity in Topological Insulator $Sb_2Te_3$ Induced by Pressure. *Sci. Rep.* **3**, 2016 (2013).

30.    Aggarwal, L., Gaurav, A., Thakur, G. S., Haque, Z., Ganguli, A. K., Sheet, G. Unconventional superconductivity at mesoscopic point contacts on the 3D Dirac semimetal $Cd_3As_2$. *Nat. Mater.* **15**, 32-37 (2016).

31.    Wang, H., *et al.* Observation of superconductivity induced by a point contact on 3D Dirac semimetal $Cd_3As_2$ crystals. *Nat. Mater.* **15**, 38-42 (2016).

32.    He, L., Jia, Y., Zhang, S., Hong, X., Jin, C., Li, S. Pressure-induced superconductivity in the three-dimensional topological Dirac semimetal $Cd_3As_2$. *Npj Quantum Mater.* **1**, 16014 (2016).

33.    Zhou, Y. H., *et al.* Pressure-induced semimetal to superconductor transition in a three-dimensional topological material $ZrTe_5$. *Proc. Natl. Acad. Sci. U.S.A.* **113**, 2904-2909 (2015).

34. Qi, Y., *et al.* Pressure-driven superconductivity in the transition-metal pentatelluride HfTe$_5$. *Phys. Rev. B* **94**, 054517 (2016).

35. Wang H., *et al.* Tip induced unconventional superconductivity on Weyl semimetal TaAs. *arXiv*, 1607.00513 (2016).

36. Li Y., *et al.* Superconductivity Induced by High Pressure in Weyl Semimetal TaP. *arXiv*, 1611.02548 (2016).

37. Kang, D., *et al.* Superconductivity emerging from a suppressed large magnetoresistant state in tungsten ditelluride. *Nat. Commun.* **6**, 7804 (2015).

38. Pan, X.-C., *et al.* Pressure-driven dome-shaped superconductivity and electronic structural evolution in tungsten ditelluride. *Nat. Commun.* **6**, 7805 (2015).

39. Qi, Y., *et al.* Superconductivity in Weyl semimetal candidate MoTe$_2$. *Nat. Commun.* **7**, 11038 (2016).

40. Autes, G., *et al.* A novel quasi-one-dimensional topological insulator in bismuth iodide [beta]-Bi$_4$I$_4$. *Nat. Mater.* **15**, 154-158 (2016).

41. Filatova, T. G. Electronic structure, galvanomagnetic and magnetic properties of the bismuth subhalides Bi$_4$I$_4$ and Bi$_4$Br$_4$. *J. Solid State Chem.* **180**, 1103-1109 (2007).

42. Werthamer, N. R., Helfand, E., Hohenberg, P. C. Temperature and Purity Dependence of the Superconducting Critical Field, $H_{c2}$. III. Electron Spin and Spin-Orbit Effects. *Phys. Rev.* **147**, 295-302 (1966).

43. Qi Yanpeng, S. W., Naumov Pavel G., Kumar Nitesh, Sankar Raman, Schnelle Walter, Shekhar Chandra, Chou F. C., Felser Claudia, Yan Binghai, Medvedev Sergey A. Topological quantum phase transition and superconductivity induced by pressure in the bismuth tellurohalide BiTeI. *arXiv*, 1610.05364 (2016).

44. Liu, C.-C., Zhou, J.-J., Yao, Y., Zhang, F. Weak Topological Insulators and Composite Weyl Semimetals: β-Bi4$X$4 ($X=$ Br, I). *Phys. Rev. Lett.* **116**, 066801 (2016).

45.    von Schnering, H. G., von Benda, H., Kalveram, C. Wismutmonojodid BiJ, eine Verbindung mit Bi(0) und Bi(II). *Z. Anorg. Allg. Chem.* **438**, 37-52 (1978).

46.    Mao, H. K., Xu, J., Bell, P. M. Calibration of the ruby pressure gauge to 800 kbar under quasi-hydrostatic conditions. *J. Geophysical Res.: Solid Earth* **91**, 4673-4676 (1986).

47.    Kresse, G., Furthmüller, J. Efficient iterative schemes for *ab initio* total-energy calculations using a plane-wave basis set. *Phys. Rev. B* **54**, 11169-11186 (1996).

48.    Blöchl, P. E. Projector augmented-wave method. *Phys. Rev. B* **50**, 17953-17979 (1994).

49.    Kresse, G., Joubert, D. From ultrasoft pseudopotentials to the projector augmented-wave method. *Phys. Rev. B* **59**, 1758-1775 (1999).

50.    Perdew, J. P., Burke, K., Ernzerhof, M. Generalized Gradient Approximation Made Simple. *Phys. Rev. Lett.* **77**, 3865-3868 (1996).

51.    Grimme, S., Antony, J., Ehrlich, S., Krieg, H. A consistent and accurate ab initio parametrization of density functional dispersion correction (DFT-D) for the 94 elements H-Pu. *J. Chem. Phys.* **132**, 154104 (2010).

52.    Grimme, S., Ehrlich, S., Goerigk, L. Effect of the damping function in dispersion corrected density functional theory. *J. Comput. Chem.* **32**, 1456-1465 (2011).

53.    Heyd, J., Scuseria, G. E., Ernzerhof, M. Hybrid functionals based on a screened Coulomb potential. *J. Chem. Phys.* **118**, 8207-8215 (2003).

54.    Heyd, J., Scuseria, G. E. Assessment and validation of a screened Coulomb hybrid density functional. *J. Chem. Phys.* **120**, 7274-7280 (2004).

55.    Mostofi, A. A. Wannier90: A tool for obtaining maximally-localised Wannier functions. *Comput. Phys. Commun.* **178**, 685-699 (2008).

56.     Fu, L., Kane, C. L. Topological insulators with inversion symmetry. *Phys. Rev. B* **76**, 045302 (2007).

57.     Murnaghan, F. D. The Compressibility of Media under Extreme Pressures. *Proc. Natl. Acad. Sci. U.S.A.* **30**, 244-247 (1944).

58.     Togo, A., Oba, F., Tanaka, I. First-principles calculations of the ferroelastic transition between rutile-type and $CaCl_2$-type $SiO_2$ at high pressures. *Phys. Rev. B* **78**, 134106 (2008).